\documentclass[final,numberedheadings]
 {aipproc}

\layoutstyle{6x9}

\begin{document}

\title{MS, S and C Stars in the Infrared. Luminosities and Mass Loss Rates.}



\author{R. Guandalini}{address={Department of Physics, University of Perugia,
Via A. Pascoli, 06123 Perugia, Italy}}

\begin{abstract}
In this note I present an outline of infrared (IR) photometric
AGB properties, based on two samples of Galactic Long Period
Variables (C- and S-type respectively). I show the various
selection criteria used during the choice of the sources and
describe the motivations of observing them at near- and mid-IR
wavelengths. I discuss the problems encountered in estimating
their luminosity and distance and motivate the methods I choose
for this purpose. Properties of the luminosity functions and of
the Hertzsprung-Russell (HR) diagrams obtained from the analysis
are discussed. Finally, the choices made for estimating of the
mass loss rates are described and preliminary results concerning
them are shown.
\end{abstract}

\maketitle

\section{Introduction}

The Asymptotic Giant Branch (hereafter AGB) phase lies at the end
of the active life for Low- and Intermediate-Mass Stars (mass
between $0.8 M_{\odot}$ and $8 M_{\odot}$). In this stage their
energy is released by two shells (burning H and He, respectively)
that ignite alternatively. For more details about nucleosynthesis
and the physical properties inside AGB stars see e.g.
\cite{busso99,nollett03,wasserburg07} and references therein.
Unfortunately, quantitative knowledge of crucial chemical and
physical parameters of AGB sources is still poor; among the
uncertain issues we emphasize particularly the mass loss rate and
the estimate of stellar luminosity, which are hampered by the
difficulties in measuring the distances for these dust-enshrouded
objects.

We are currently performing an analysis of Galactic AGB stars of
different chemical composition, looking for correlations between
their basic parameters (bolometric luminosity, mass loss rate,
photometric colors \dots ). Our main aim is that of establishing
observationally-based criteria permitting a more quantitative
determination of mass loss rates and luminosities for the various
types of AGB stars thus providing general rules that could help
in the improvement of stellar codes.

Extensive samples of Galactic AGB stars (M, S-type, C-rich) have
been collected. They must be large enough so that conclusions on
them have a good statistical significance; they also need to have
detailed and accurate Spectral Energy Distributions (SEDs) at
near- and mid-IR wavelengths and, whenever possible, reliable
measurements of mass loss rates and distances. Results of this
research have been already published for carbon-rich stars
(\cite{gua06}, hereafter Paper I), and are in press for S-type
stars (\cite{guabus08}, hereafter Paper II).

In the following section (Section 2) we present the two samples
analyzed in Paper I and Paper II, explaining in detail through
which criteria they have been divided into the different
sub-samples. In Section 3 we show the importance of performing
observations of AGB stars at mid-IR wavelengths (at least up to 20
$\mu$m). The methods adopted to estimate the luminosity and/or
distance are described in Section 4. In Section 5 we discuss the
results on luminosity functions and HR diagrams (in particular
for S-type stars). In Section 6 we comment briefly on our choices
to estimate mass loss rates and show the first results for S-type
stars from our research. Finally, in Section 7 preliminary
conclusions are derived.

\section{Samples of AGB Stars}

Two samples of AGB stars have been created to perform this
analysis: one of carbon-rich sources (as discussed in Paper I)
and the second of O-rich but s-element enhanced sources (S-type
stars, as described in Paper II). Both samples have photometric
observations available at mid-IR wavelengths (at least in the 8
$\mu$m region). They are subsequently organized in sub-samples
according to the quality and extension of the database available.

\subsection{Carbon-rich Stars}

AGB sources that have been considered in this sample are
chemically carbon-rich. Moreover, they have at least one
published estimate of the mass loss rate, done according to the
criteria we consider most suitable for the purpose (see Section 6
for details).

We have divided the carbon stars into three sub-samples according
to the available knowledge of their variability and distance:
\begin{itemize}
\item \textbf{Sub-sample 1} (65 sources) is made out of our "best"
sources. They are characterized by reliable estimates of the
distance, obtained through recent updates of original Hipparcos
measurements \cite{berg2002,berg2005}. Moreover these sources are
all of a known variability type.

\item \textbf{Sub-sample 2} (50 sources). Here the estimates of
the distances derive from various methods (mainly from the
application of the period-luminosity relations valid for C-rich
stars). All the stars of this sub-sample are again of known
variability types.

\item \textbf{Sub-sample 3} (150 sources). These sources are the
less studied of the whole sample. Their distances are computed
through various methods, as found in the literature, the most
commonly used being the one from \cite{groe2002}. They are all of
"unknown" variability types.
\end{itemize}

For a more extensive discussion of this sample of Galactic C-rich
AGB stars the reader is referred to Paper I.

\subsection{S-type Stars}

In order to prepare the sample under analysis for the S-type
Galactic stars, we started from the extended lists by
\cite{steph1984,steph1990}, containing O-rich evolved red giants
with known or suspected "S star-like" chemical peculiarities. We
selected a total of $\simeq$600 sources for which measurements in
the near-IR (from 2MASS) and in the mid-IR (from ISO, MSX or
IRAS--LRS) are available. The chosen sources fall again into
different categories, depending on the information we could
collect on their IR colors, distance, variability type and period.

The S-type stars have therefore been organized in four
sub-samples:
\begin{itemize}
\item \textbf{Sub-sample \emph{A}} (21 sources): stars with
ISO-SWS observations at mid-IR wavelengths up to 45 $\mu$m. They
are fundamental for the calculation of bolometric corrections.

\item \textbf{Sub-sample \emph{B}} (26 sources). These sources
have mid-IR measurements from MSX and/or IRAS--LRS (not from
ISO--SWS). Their distances are estimated astrometrically from
Hipparcos measurements.

\item \textbf{Sub-sample \emph{C}} (40 sources). This class
contains only stars of Mira-type variability with reliable studies
of their light-curves and therefore, good estimates of their
period. Photometric observations in the mid-IR have been performed
by MSX or IRAS--LRS. Distances are estimated through
period-luminosity relations valid for O-rich stars (see Section
4).

\item \textbf{Sub-sample \emph{D}} ($\simeq$500 sources). This
class includes all the other sources. They have "good" photometric
observations at mid-IR wavelengths (at least 8.8 $\mu$m) but we
could not find estimates of the distance.
\end{itemize}

The four sub-samples are divided in further sub-categories
according to:
\begin{enumerate}
\item the spectral type (usually S, MS, (M)(S)\footnote{(M)(S)
indicates an uncertain classification between the two classes
above}, SC, plus unknown sources and anomalous cases);

\item the variability type (Mira, Semiregular,
Irregular; unknown sources exist also in this case).
\end{enumerate}

A detailed analysis of this large sample of S-type stars is
presented in Paper II.

\section{The importance of IR for AGB stars}

%
%
\begin{figure}[tb!]
\includegraphics[height=.355\textheight,clip=true]{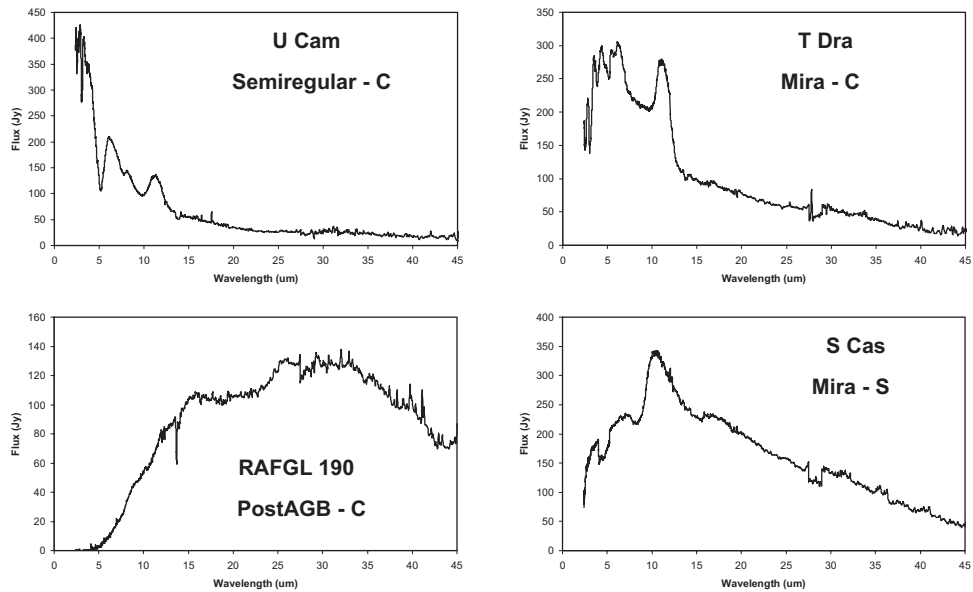}
\caption{ISO-SWS spectra for a C-rich Semiregular variable (U Cam,
top left), a C-rich Mira variable (T Dra, top right), a C-rich
post-AGB star (RAFGL 190, bottom left) and S-type Mira variable (S
Cas, bottom right).}\label{fig1}
\end{figure}

Evolution during the AGB phase has many "much-debated" topics:
their study could be greatly helped by photometric observations at
IR wavelengths.

In particular, an extended wavelength coverage is fundamental in
the exam of the SEDs of AGB stars. Indeed, Fig. \ref{fig1} shows
that evolved AGB stars (Mira variables and post-AGBs) emit large
part of their flux at mid-IR wavelengths. This fact is
particularly evident for C-rich sources, to the point that mid-IR
observations are necessary to obtain good estimates of bolometric
luminosity, that otherwise would be significantly underestimated.

AGB stars are characterized by strong variability at optical
wavelengths. Variability of the flux at IR wavelengths is less
relevant if compared with the optical region, because in IR the
contribution from the dynamical photosphere is less relevant.
However, IR variability is not negligible. Indeed, the few Mira
sources that have been observed several times by ISO with the SWS
spectrometer show remarkable fluctuations in the mid-IR region.
By contrast, SR variables do not show any appreciable variation of
the flux (see \cite{busso7} and its figures for more details).
There is not yet an agreement on the origin and properties of
this phenomenon: it needs to be examined in more detail in the
mid-IR to understand its nature.

\section{Luminosity and Distance}

The SEDs from ISO-SWS available for the two samples of AGBs plus
the near-IR photometric points (from 2MASS) can be integrated in
order to infer a reliable (apparent) bolometric magnitude, making
use of the relations of fundamental photometry (Glass,
\cite{glass}). If we consider the S-type sample, the sources of
the sub-sample \emph{A} give us the tools for computing the
bolometric corrections, which turn out to be tightly correlated
with near-to-mid infrared colors and can be used for deriving
accurate bolometric magnitudes for the stars of sub-sample
\emph{B}, where the ISO-SWS measurements are not available (see
Paper II for details). The same approach was adopted in Paper I in
analyzing C star luminosities.

With the procedure described above, we can compute the bolometric
magnitude whenever a source has observations in at least one of
the relevant IR colors.

S-type stars that belong to the sub-sample \emph{C} offer the
possibility of an independent way for estimating luminosities. In
fact, their Mira variability, of known period, allows for the
application of period-luminosity relations, yielding directly the
absolute magnitudes.

Concerning  period-luminosity relations, for Galactic C-rich Miras
they have been computed by Whitelock et al. (see \cite{white2006}
and references therein) and summarized in the relation:
\begin{equation}
M_{bol} = -2.54 log(P) + 1.87
\end{equation}
where $P$ is the variability period of the Mira and $M_{bol}$ is
its absolute luminosity. Instead, searching a similar
period-luminosity relation for the O-rich Miras (and also S-type
stars) is definitely more difficult. In the literature there is no
affordable, unique estimate of this relation. We choose (mainly
for the sake of homogeneity) to follow the indications from Feast
and Whitelock (see \cite{feast2004} and references therein for a
general review on this topic). Obviously the P-L relation for
O-rich Miras (and S-type stars) will be different from the one
obtained for C-rich Miras. Our choice fell on two different P-L
relations obtained using period correlations either to the
absolute bolometric luminosity (\cite{white94}, Eq. 2) or to the
absolute K-band luminosity (\cite{feast2004}, Eq. 3):
\begin{equation}
M_{bol} = -3.00 log(P) + 2.80
\end{equation}

\begin{equation}
M_{K} = -3.47 log(P) + 1.00
\end{equation}

We made use of both methods, because they have both advantages
and disadvantages. In principle Eq. 2 (from \cite{white94}) would
be probably the best relation available, as it involves $M_{bol}$
directly. However, its last version is quite "old" (1994,
\cite{white94}). Instead Eq. 3 (from \cite{feast2004}) has been
more recently updated and extensively examined \cite{feast2004}
but gives us "only" an estimate of the absolute magnitude in the
K--band. We adopted both relations to calculate two independent
estimates of distance:
\begin{itemize}
    \item First case -- using Eq. 2. Here we find an estimate of
    the apparent bolometric  magnitude through the bolometric
    corrections obtained in Paper II, then we compare it with the
    absolute bolometric magnitude from Eq. 2 to estimate the
    distance of the S Mira.
    \item Second case -- using Eq. 3. Here we apply a similar
    procedure comparing the apparent K--magnitude from 2MASS with
    the absolute K--magnitude from Eq. 3 to obtain another
    estimate of the S Mira's distance.
\end{itemize}
Finally, we make an average between the two distance estimates and
with the final result we calculate the absolute magnitude of the
star applying  the "final" distance to the apparent bolometric
magnitude obtained form our bolometric corrections. See Paper II
for a more complete discussion of this topic.

We are trying to improve our knowledge on the variability
properties of Mira and Semiregular AGB stars. One of the analyses
that we are performing is centered on the re-observation of
Maffei's stars with Small--IRAIT, in order to confirm the periods
and light--curves obtained by Maffei and Tosti (the catalogue is
available online at
\emph{http://astro.fisica.unipg.it/atlasmaffei/main.htm}). See the
contribution from Briguglio in this conference for more details
\cite{brigu}.

\section{Luminosity Functions and HR Diagrams}

Here we present few results obtained using the bolometric
magnitudes computed through the methods presented in the previous
section. We do not discuss in detail the luminosity function for
Carbon stars because it has been already presented in Paper I.
Here we remark only that, according to our data, Galactic C-rich
AGB stars show luminosities in the range $-4.5 / -6$. This is in
agreement with the theoretical predictions from the FRANEC model
(see \cite{stran} and the contribution from Cristallo, this
conference). We don't find any problem of underluminous C-rich
stars, contrary to suggestions in the past (\cite{costafrogel},
see also Paper I)

%
%
\begin{figure}[tb!]
\includegraphics[height=.245\textheight,clip=true]{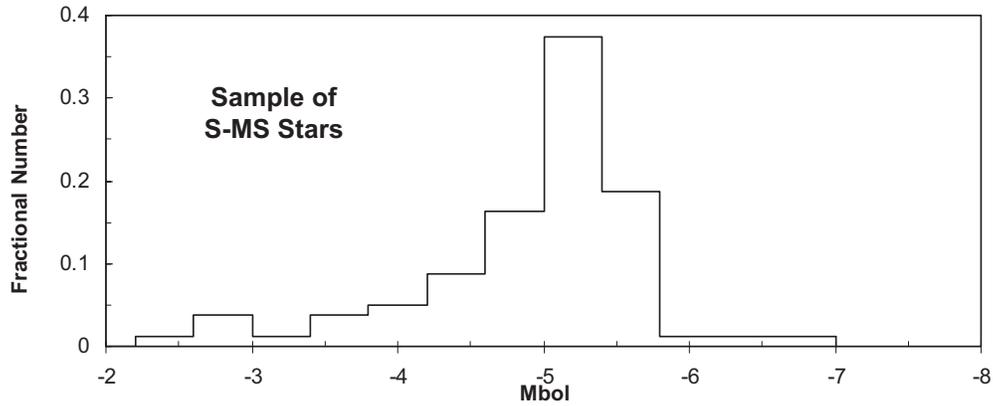}
\caption{The histogram showing the fractional number of S-type
stars per magnitude interval.}\label{fig2}
\end{figure}

%
%
\begin{figure}[tb!]
\includegraphics[height=.225\textheight,clip=true]{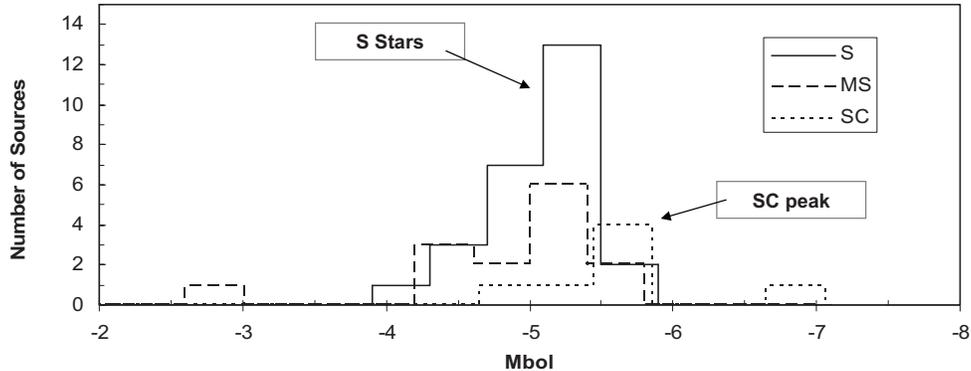}
\caption{Luminosity Functions from the same sample already shown
in Fig. \ref{fig2}. The original data have been divided into 3
sub-samples, according to the source spectral type
(S--MS--SC).}\label{fig3}
\end{figure}

%
%
\begin{figure}[tb!]
\includegraphics[height=.225\textheight,clip=true]{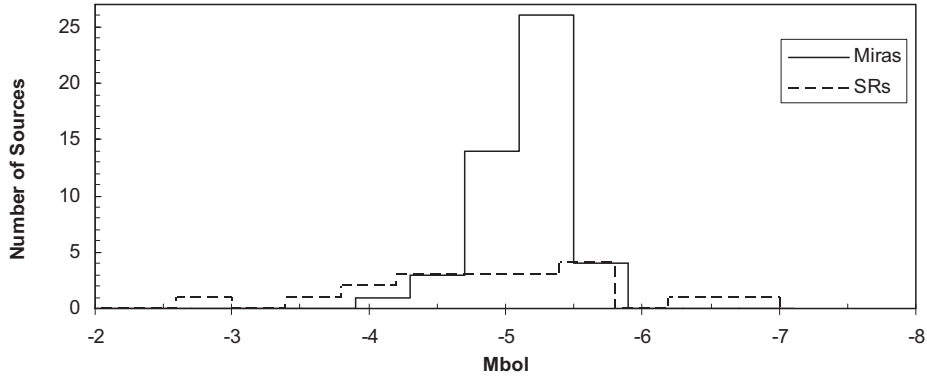}
\caption{Luminosity Functions from the same sample already used
for Fig. \ref{fig2}. We present two sub-samples formed according
to the variability type (Mira--Semiregular).}\label{fig4}
\end{figure}

In Fig. \ref{fig2} the luminosity function for the S-type stars
from sub-samples \emph{A}--\emph{B}--\emph{C} is shown. The spread
in the luminosities for these 80 sources is again in agreement
with the FRANEC models. The luminosities are in the range $-4.2 /
-5.7$ while the main peak is between -4.9 and -5.2: Galactic S
stars seem to be on average only slightly less luminous of the
C-rich ones. At this stage we cannot say whether this apparent
difference is statistically significant: in fact, the two
Luminosity Functions are overlapping. It is clear that attributing
the spectral distinction among the classes  MS--S--SC--C to a
simple evolutionary sequence, is not sufficient.

We divide the sample used for the luminosity function of Fig.
\ref{fig2} in three sub-groups according to the different spectral
types (S--MS--SC). In this way we obtain three luminosity
functions that are shown in Fig. \ref{fig3}. The S stars present a
peak in their distribution in the range $-4.7 / -5.5$, the SC
stars occupy the high luminosity tail of the distribution with a
peak at magnitudes around -5.6. Instead, the MS sample is
overlapping both the S and SC sub-groups without a well-defined
peak.

Figure \ref{fig4} shows other luminosity functions. However in
this case the sample of S-type stars has been divided according to
the variability type of the sources (Mira or Semiregular). The
Mira's Luminosity Function presents a well defined peak and cover
mainly the range between -4.8 and -5.5. On the other hand, the
distribution of Semiregulars is more or less uniform over all the
range of luminosities.

%
%
\begin{figure}[tb!]
\includegraphics[height=.389\textheight,clip=true]{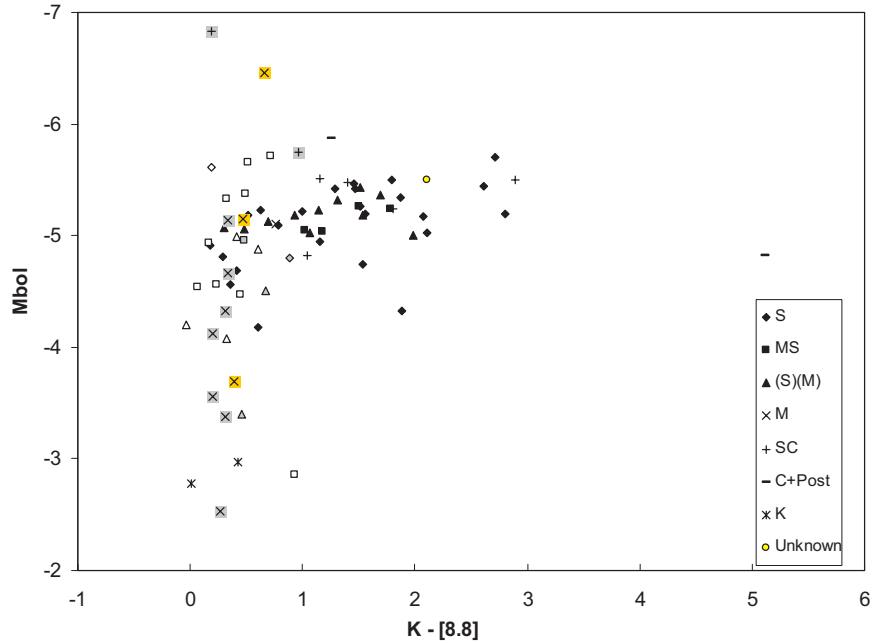}
\caption{The HR diagram for the sample of S stars adopting the
near-to-mid-IR colour K--[8.8] as abscissa. The sources are
divided into several sub-groups according to their spectral type,
as shown in the legend inside the plot. For each sub-group, filled
points are Mira variables, while blank points are Semiregular or
Irregular ones. }\label{fig5}
\end{figure}

Finally, we present in Fig. \ref{fig5} a preliminary result on
the S-star HR diagram, adopting a near-to-mid-IR color as the
abscissa. We refer to Paper II for a more complete discussion.
Here we emphasize a few main features after dividing the points
according to the spectral classification and the variability
types. We note that semiregular S stars are essentially
unreddened, while the Miras do show some reddening. However, if we
compare S stars with the Carbon-rich sample (see Paper I), C stars
are in general more deeply enshrouded by dust.

\section{Mass Loss}

As a last topic of this contribution we present a rough analysis
of mass loss rates from AGB stars including a preliminary result
for the S-type stars. For C-stars see Paper I.

Longstanding efforts have been devoted to describe the mass loss
mechanisms, either with phenomenological models or with
sophisticated hydrodynamical approaches
\cite{salpeter,knappmorris,winters03,wachter02}. Despite this, our
quantitative knowledge of AGB winds is still poor. Our choice has
been that of using mass loss rates available form the literature.
We considered only those computed through observations at
radio-wavelengths of the CO lines: this type of approach is known
to give an estimate of the "average" rate of the AGB source's
stellar wind, integrated over a long time. We upgraded the
estimates of the mass loss rate adopting updated measurements of
the distance or, in the case of the S-type sample (see Paper II)
giving directly our estimate of the distance through the use of
the Period-Luminosity relations (see also Section 4).

The approaches to obtain mass loss estimates through the
observations of CO lines that are available in the literature are
several. In particular we adopt:
\begin{enumerate}
    \item \emph{First Choice} (our best choice): The analysis performed by the Stockolm's group:
    we refer to  Schoier and Olofsson \cite{schoier} for C-rich stars, Olofsson
    et al. \cite{olof} for M stars and Ramstedt et al. \cite{ramstedt} for S-type stars.
    \item \emph{Second Choice}: the method used by Loup et al. \cite{loup} and Winters et al.
    \cite{winters03}. In this case the mass loss rate and the size of the CO
    envelope are both estimated in a bi-parametric system.
    \item \emph{Third Choice}: the method used by Groenewegen et al.
    (\cite{groe1998} for S stars, \cite{groe2002} for C-rich stars).
    Here a constant size for the CO envelope is adopted for all the sources.
    However, this method is more reliable for stars with an optically thick CO emission.
    Due to this fact this method can sometime underestimate the wind
    efficiency by factors 3--4 (as said by \cite{schoier}).
\end{enumerate}
The analysis performed on the carbon-rich sample is shown in Paper
I.

\begin{figure}[tb!]
\includegraphics[height=.279\textheight,clip=true]{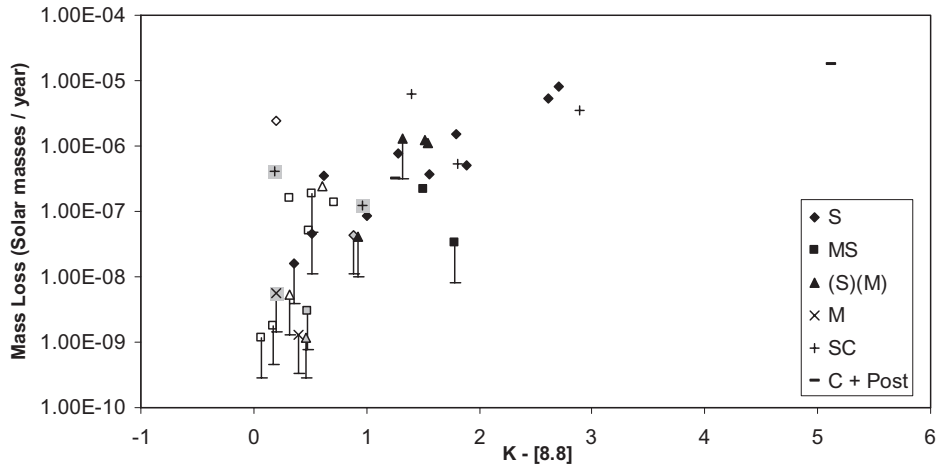}
\caption{Relation linking the mass loss rates of S-type stars to
the near-to-mid-IR colour K--[8.8]. The sources are divided into
several sub-groups according to their spectral types as shown in
the legend inside the plot. For every sub-group, filled points are
Mira variables, while blank points are Semiregular or Irregular
ones. The data with the downward-pointing error bars are upper
limits.}\label{fig6}
\end{figure}

Figure \ref{fig6} displays a preliminary result obtained plotting
the mass loss rates against a near-to-mid-IR color (the same one
already used as abscissa of Fig. \ref{fig5}) in the case of the
sample of S-type stars. The sources show a behaviour similar to
the one already analyzed in the C-rich sample (see Paper I); here
we remark a few main properties: 1) Miras have stronger mass loss
rates and are redder if compared with Semiregulars and Irregulars;
2) a relation of the same type as found for C stars exists for
S-stars suggesting the association of strong mass loss with
strong IR excess for AGB stars of any type, as expected is the
wind is driven by radiation pressure on dust grains; 3) this
relation, if real, is however very scattered for both samples; 4)
van Loon et al. \cite{vanloon2005} have suggested that the mass
loss rate should be expressed as a function of both the bolometric
luminosity and the stellar effective temperature (photometric
color); 5) we remind that mass loss processes in AGB stars are
surely not "simple": the large variations of the mass loss rates,
due to different causes, that occur during the AGB phase
certainly contribute largely to the scattering that we see in
Fig. \ref{fig6}. Finally, we remark that there is no clear
distinction in the mass loss rates of S-type and C-rich stars,
but carbon stars become much redder by a large factor (see Paper
I and Paper II).

\section{Conclusions}

%
%

We have analysed two sample of Galactic AGB stars: the first is
made of carbon rich sources and the second one of S-type stars. We
looked for methods to obtain reliable estimates of luminosities
and distances (bolometric corrections and Period--luminosity
relations). The problem of the large IR excess typical of evolved
AGBs has been shown together with its implications in the
estimates of the stellar luminosity.

C- and S-stars Luminosity Functions are in the ranges -4.5/-6 and
-4.2/-5.7 respectively (in agreement with expectations from the
FRANEC model). The Luminosity Functions of the two samples are
overlapping (with S stars being slightly less luminous). It is
clear that attributing the spectral distinction among the classes
MS--S--SC--C to a simple evolutionary sequence, is not sufficient.

SC stars occupy the high luminosity tail of the S-type stars
distribution. It is highly uncertain if this is due to
evolutionary properties. Semiregular S stars are essentially
unreddened and S stars in general are less reddened than C stars.

There is no clear distinction between S and C classes in what
concerns the relation of mass loss rates to near-to-mid-IR colors,
but C stars become much redder. High-reddened C stars (evolved
Miras and post-AGBs) possibly lose more mass than S stars but the
scatter and the uncertainties of the rates are large.

\begin{theacknowledgments}
R.G. acknowledges support in Italy by MIUR under contract
PRIN2006-022731 and the University of Perugia for a postdoctoral
fellowship.
\end{theacknowledgments}

\bibliographystyle{aipproc}                           

\end{document}